\documentclass[twocolumn]{aastex631}
\usepackage{amsmath}
\usepackage{comment}
\usepackage{natbib}
\usepackage{booktabs, multirow}
\usepackage{changepage,threeparttable}
\usepackage{appendix}

\shortauthors{Moharana, Hema \& Pandey}
\shorttitle{Solar Helium Abundance}

\begin{document}

\title{Helium Abundance of the Sun: A Spectroscopic Analysis}

\author[0009-0007-9411-0284] {Satyajeet Moharana}
\affiliation{Indian Institute of Science Education and Research, Berhampur, 760003 India; \href{mailto:satyajeetm19@iiserbpr.ac.in}{satyajeetm19@iiserbpr.ac.in}}
\affiliation{Indian Institute of Astrophysics; Bengaluru, 560034 India; \href{mailto:hema.bp@iiap.res.in}{hema.bp@iiap.res.in}, \href{mailto:pandey@iiap.res.in}{pandey@iiap.res.in}}

\author[0000-0002-0160-934X] {B. P. Hema}
\affiliation{Indian Institute of Astrophysics; Bengaluru, 560034 India; \href{mailto:hema.bp@iiap.res.in}{hema.bp@iiap.res.in}, \href{mailto:pandey@iiap.res.in}{pandey@iiap.res.in}}
\author[0000-0001-5812-1516] {Gajendra Pandey}
\affiliation{Indian Institute of Astrophysics; Bengaluru, 560034 India; \href{mailto:hema.bp@iiap.res.in}{hema.bp@iiap.res.in}, \href{mailto:pandey@iiap.res.in}{pandey@iiap.res.in}}

\begin{abstract}
Determining the He/H ratio in cool stars presents a fundamental astrophysical challenge. While this ratio is established for hot O and B stars, its extrapolation to cool stars remains uncertain due to the absence of helium lines in their observed spectra. We address this knowledge gap by focusing on the Sun as a representative cool star. We conduct spectroscopic analyses of the observed solar photospheric lines by utilizing a combination of MgH molecular lines and neutral Mg atomic lines including yet another combination of CH and C$_{2}$ molecular lines with neutral C atomic lines. Our spectroscopic analyses were further exploited by adopting solar model atmospheres constructed for distinct He/H ratios to determine the solar photospheric helium abundance. The helium abundance is determined by enforcing the fact that for an adopted model atmosphere with an appropriate He/H ratio, the derived Mg abundance from the neutral Mg atomic lines and that from the MgH molecular lines must be the same. Ditto holds for the C abundance derived from neutral C atomic lines and that from CH lines of the CH molecular band and C$_{2}$ lines from the C$_{2}$ Swan band. The estimated He/H ratio for the Sun is discussed based on the one-dimensional local thermodynamic equilibrium (1D LTE) model atmosphere. The helium abundance (He/H = 0.091 $^{+ 0.019}_{- 0.014}$) obtained for the Sun serves as a critical reference point to characterize the He/H ratio of cool stars across the range in their effective temperature. Using this derived He/H ratio, the solar mass fractions are determined to be $X_{\odot}$ = 0.7232 $^{+0.0305}_{-0.0377}$, $Y_{\odot}$ = 0.2633 $^{+0.0384}_{-0.0311}$, and $Z_{\odot}$ = 0.0135 $^{+0.0006}_{-0.0007}$.
\end{abstract}

\clearpage


\section{Introduction} \label{sec:intro}
Fundamentally, elemental abundances of all astrophysical entities are compared against their solar values. This makes the chemical composition of the Sun a benchmark and an essential reference in the field of astronomy and astrophysics including cosmology, astroparticle, space and geophysics. Over a century, advances have been made in characterizing the complete solar composition from the significant studies of \cite{1929ApJ....70...11R, RevModPhys.28.53, 1960ApJS....5....1G, 1968MNRAS.138..143L, 1978MNRAS.182..249L, 1989GeCoA..53..197A, 1998SSRv...85..161G} to the more recent studies of \cite{2003ApJ...591.1220L, 2005ASPC..336...25A, 2009ARA&A..47..481A, 2011SoPh..268..255C, 2021A&A...653A.141A}. In this context, it is worth noting \cite{2020JApA...41...41A} that gives an overview of the advances and the way forward in spectroscopic analysis.

\par However, spectroscopic determination of helium abundance, \textit{i.e.}, $\log \epsilon$(He), or the helium-to-hydrogen (He/H) ratio, in the solar photosphere has always remained a fundamental astrophysical challenge due to the absence of helium line transitions in the photospheric absorption spectrum of the Sun. Though measurement of solar helium abundance can be obtained from observing coronal sources, including the solar cosmic rays \citep{1967Natur.215...43L}, solar wind \citep{1969SoPh....8..435O}, and solar energetic particles \citep{2021LNP...978.....R} or from the chromospheric line intensities \citep{1973RvGSP..11..115H}, but these measurements do not essentially demonstrate the photospheric helium content, for example, one possible reason may be due to the FIP effect as discussed in \cite{2015LRSP...12....2L}. 

\par \cite{1989GeCoA..53..197A} derived the proto-solar helium content ($\log \epsilon$(He) = 10.99) from H\,{\sc ii} regions and B-type stars which share similar metallicity as that of the Sun. This determination led to the adoption of a He/H ratio of 0.1 ($\log \epsilon$(He) = 11.00) for model atmospheres used in solar abundance analysis over the years, under the assumption that this ratio remains consistent across both hot and cool stars. The adopted helium abundance is in good agreement with that of a recent study by \cite{2012A&A...539A.143N} for early B-stars.

\par Solar helium abundance have been estimated from indirect methods that is one through helioseismology that determines the He/H ratio accurately in the solar convection zone by analysing the second ionisation region of Helium \citep{2004ApJ...606L..85B, 2005MNRAS.361.1187M, 2007MNRAS.375..861H}. This method, however, is sensitive towards the adopted equation of state \citep{2008PhR...457..217B} and the assumed metallicity of the reference solar model. The important problem is that the predictions of the standard solar model, for the adopted downward revised solar abundances of \cite{2009ARA&A..47..481A}, do not agree with the helioseismic determinations of the sound speed, the depth of the convection zone and the abundance of helium in this layer.
 
\par In this study we have adopted a novel technique, similar to that described by \cite{2020ApJ...897...32H} for cool giants, to spectroscopically determine the solar photospheric He/H ratio. This new method was the outcome of our earlier two studies: \cite{2014ApJ...792L..28H, 2018ApJ...864..121H}. In the following Sections, we describe the solar spectrum, the adopted model atmospheres, and the abundance analyses procedure.

\section{The Solar Spectrum} \label{sec:spectrum}
For this study, we have used a high-resolution, high signal-to-noise ratio solar spectrum and is from the National Solar Observatory (NSO) archives. This solar flux spectrum, as documented by \cite{1984sfat.book.....K}, was observed using the McMath solar telescope equipped with a Fourier Transform Spectrometer (FTS). The spectrum has a resolving power $R$ $(\lambda/\Delta\lambda)$ $\sim$400,000 and a signal-to-noise ratio of about 1000 per pixel in the wavelength range 3400-9300 {\AA}. The observation involved directing unfocused Sunlight from the solar heliostat into the FTS instrument. This method captures the solar disk in its entirety, effectively representing the Sun as a star in our observations \citep{2000vnia.book.....H}.

\par This FTS solar spectrum was used by \cite{1998A&AS..131..431A} to compile a precise wavelength catalogue in the optical spectrum of the Sun. Additionally, the equivalent widths measured from this spectrum are in excellent agreement with other solar spectrum studies that have been referred to in this paper.

\section{Abundance analysis} \label{sec:abundance}
The observed solar spectrum, as discussed above, is analyzed in local thermodynamic equilibrium (LTE) using a radiative transfer code MOOG \citep{2012ascl.soft02009S} combined with a star's model atmosphere to compute the absorption spectrum or to predict the equivalent width of an absorption line. In this study we have adopted ATLAS12 model atmospheres with different He/H ratios. These model atmospheres were computed based on the plane-parallel and LTE approximation by taking into account the line-blanketing effect, see \cite{2014dapb.book...39K} and \cite{2020ApJ...897...32H} for details.

\par To adopt a model atmosphere computed for the normal He/H ratio of 0.1, the input abundances of H and He required by MOOG are $\log \epsilon$(H) = 12.00 and $\log \epsilon$(He) = 11.00. Similarly, for a model atmosphere computed for a He/H ratio of 0.125, the input abundances of H and He, which need to be provided to MOOG, are $\log \epsilon$(H) = 11.974 and $\log \epsilon$(He) = 11.071. These input abundances of H and He for different He/H ratios are calculated by utilizing a standard normalization relation:  \[ \sum_{i} \mu_{i}E_{i} = \mu_{\rm H}H + \mu_{\rm He}He + \sum_{i\,=\,3} \mu_{i}E_{i} = 10^{12.15} \] where $\mu_{i}E_{i}$ represents the total mass of an element E, having atomic number $i$ present in the stellar photosphere, with $\mu_{i}$ and $E_{i}$ denoting the atomic mass and abundance by number for the element E, respectively. Assuming that H and He are the primary components of the stellar photosphere, while all other elements present are in trace amounts, \textit{i.e.}, \[ \sum_{i\,=\,3} \mu_{i}E_{i} \rightarrow 0 \] then, $H + 4He = 10^{12.15}$, since $\mu_{\rm H} = 1$, and $\mu_{\rm He} = 4$. Note that, conventionally, log of $H$ is $\log \epsilon$(H) and log of $He$ is $\log \epsilon$(He) or, in general, log of $E$ is $\log \epsilon$(E).

\subsection{Adopted Solar Parameters} \label{sec:parameters}
\par The solar parameters, such as, effective temperature $(T_{\rm eff})$, surface gravity (log \textit{g}) and metallicity ([Fe/H]) were adopted from \cite{2021A&A...653A.141A} and \cite{gray_2021}, as $T_{\rm eff}$ = \mbox{5773 $\pm$ 16 $\rm K$}, $\log g$ = 4.4374 $\pm$ 0.0005 (cgs), and [Fe/H] = 0.0. We have also adopted a microturbulence ($\xi_{\rm t}$) of 1 ${\rm km\,s^{-1}}$ as suggested by \cite{2021A&A...653A.141A}.

\subsection{Equivalent Width Analyses and Spectrum Syntheses}\label{sec:ew}

\par
\begin{figure}
\centering
\includegraphics[scale=0.33]{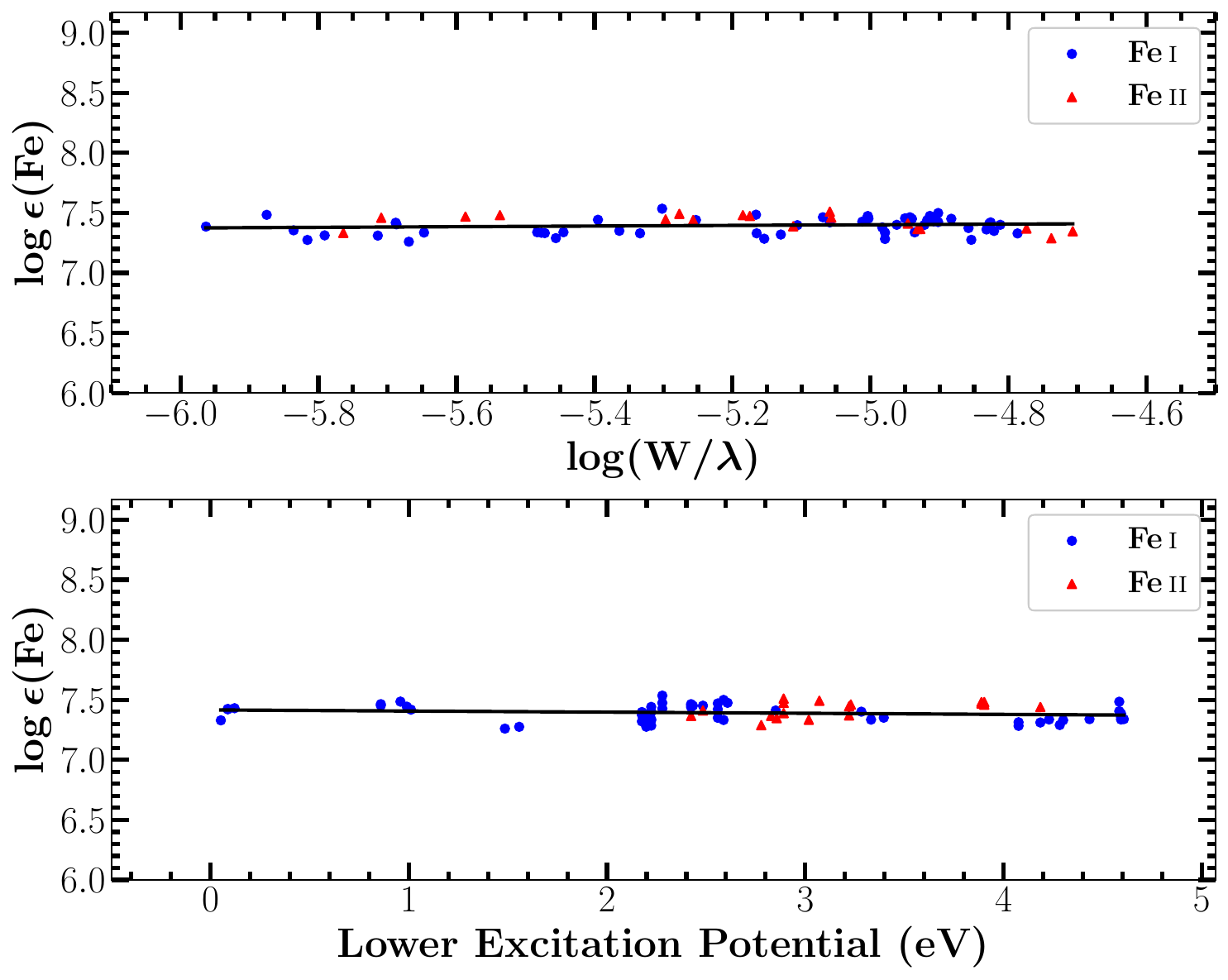}
\caption{$\log \epsilon$(Fe) versus $\log (\rm W/\lambda)$ for the adopted $\xi_{\rm t}$ = 1.0 ${\rm km\,s^{-1}}$ (top panel). $\log \epsilon$(Fe) versus lower excitation potential (LEP) for the adopted ($T_{\rm eff}$, $\log g$) = (\mbox{5773 $\rm K$}, \mbox{4.44 cgs}) (bottom panel).} \label{fig:temp_mt}
\end{figure}

 To validate the adopted solar parameters, an abundance analysis was performed on the observed solar photospheric spectrum. Neutral and singly ionized absorption lines of iron (Fe\,{\sc i} and Fe\,{\sc ii}) were used as probes to verify the excitation and the ionization balance including the adopted microturbulence. The Fe\,{\sc i} and Fe\,{\sc ii} lines are taken from \cite{2000A&A...359..743A}. Abundances of iron were derived from the measured equivalent widths of Fe\,{\sc i} and Fe\,{\sc ii} lines by using an ATLAS12 model atmosphere with He/H ratio 0.1 and the adopted solar parameters. The derived iron abundance, $\log \epsilon$(Fe), versus the line's reduced equivalent width (REW), log ($\rm W/\lambda$), and its lower excitation potential (LEP), are shown in Figure \ref{fig:temp_mt}(top panel) and \ref{fig:temp_mt}(bottom panel), respectively. Inspection of Figure \ref{fig:temp_mt}(top panel) validates the adopted microturbulence as no trend is noticed in the derived Fe abundances with respect to REW. Similarly, inspection of Figure \ref{fig:temp_mt}(bottom panel) suggests no trend in the derived Fe abundances with respect to LEP, satisfying the excitation as well as the ionization balance for the adopted effective temperature and the surface gravity. Note that, Fe\,{\sc i} and Fe\,{\sc ii} lines with a range in their LEPs return similar Fe abundances by satisfying the excitation as well as the ionization balance. Hence, the above tests confirm and validate the adopted solar parameters, without any ambiguity, for conducting the abundance analysis.

\par In this study, we have primarily focused on the absorption features of neutral atomic lines of magnesium and carbon as well as molecular lines of their compounds involving hydrogen. Several Mg\,{\sc i} atomic lines and MgH molecular lines of the \mbox{MgH $A-X$ (0, 0)} band, as well as C\,{\sc i} atomic lines including a forbidden transition that is [C\,{\sc i}] line at 8727.126{\AA}, CH molecular lines of the CH electronic ($A-X$) band, and C$_{2}$ molecular lines of the \mbox{C$_{2}$ Swan (0, 0)} band, were identified in the observed solar spectrum. These observed spectral features were accordingly subjected to equivalent width analyses and spectrum syntheses.

\par In this study, abundance analyses was conducted for the adopted solar parameters of model atmospheres having 8 different He/H ratios: 0.075, 0.085, 0.100, 0.125, 0.135, 0.150, 0.175 and 0.200.

\subsubsection{Molecular Lines}\label{sec:synthesis}
The spectrum synthesis code MOOG combined with ATLAS12 model atmospheres was used to synthesize MgH, CH, and C$_{2}$ Swan molecular lines present in the observed solar spectrum. For this purpose, the solar rotational velocity $(v\sin{i})$ and macroturbulent velocity $(\xi_{\rm T})$, and the resolution of the observed solar spectrum at a given wavelength were required.

\par The adopted values for the solar rotational velocity $(v\sin{i})$ and macroturbulent velocity $(\xi_{\rm T})$ are 1.7 ${\rm km\,s^{-1}}$ and 3.2 ${\rm km\,s^{-1}}$, respectively. These values are in fair agreement with \cite{2012MNRAS.422..542P} for $v\sin{i}$ and with \cite{2022A&A...668A...9H} for $\xi_{\rm T}$. The adopted resolution, derived from the resolving power of the observed solar spectrum, as represented by a gaussian of FWHM is 0.02{\AA} at around 6500{\AA}.

\par Our adopted values for $v\sin{i}$ and $\xi_{\rm T}$ were obtained from the equivalent width analyses and the spectrum syntheses of the observed Fe\,{\sc i} lines. The source of the required atomic data and the measured equivalent widths of these Fe\,{\sc i} lines are discussed in Sec. \ref{sec:parameters}. The measured equivalent width of an individual Fe\,{\sc i} line provides the Fe abundance for the best adopted solar model with the parameters: ($T_{\rm eff}$, $\log g$, [Fe/H], $\xi_{\rm t}$) = (\mbox{5773 $\rm K$}, \mbox{4.44 cgs}, 0.0, 1.0 ${\rm km\,s^{-1}}$). The absorption profile of the Fe\,{\sc i} line is then synthesized for the above derived Fe abundance combined with the adopted solar model. The best fit to the observed Fe\,{\sc i} line is then obtained by tuning the two parameters, $\textit{v}\sin{i}$ and $\xi_{\rm T}$. This procedure is then followed for a set of observed Fe\,{\sc i} lines to determine the mean $v\sin{i}$ and $\xi_{\rm T}$. Synthesis of Fe\,{\sc i} line at 6574.229{\AA} is shown in Figure \ref{fig:synth}(a) as an example.

\vspace{0.5cm}
\textbf{MgH:} The solar $^{24}$MgH molecular lines for the \mbox{$A-X$ (0, 0)} molecular band are from \cite{1971MNRAS.154..265L}. \cite{1971MNRAS.154..265L} note that all P branch $^{24}$MgH lines are blended with $^{25}$MgH and $^{26}$MgH lines, and in their Table 1 the Q and R branch lines that are blended with $^{25}$MgH or $^{26}$MgH features are marked with asterisk. Note that P, Q, and R branches refer to different types of ro-vibronic molecular transitions and are classified based on the initial ($J^{\prime \prime}$) and final ($J^{\prime}$) state quantum numbers of the transition. The transition lines with $\Delta J$ = -1 belong to the P branch. Similarly, $\Delta J = 0$ and $\Delta J = 1$ correspond to the Q and R branches, respectively \citep{banwell1994fundamentals}.

\par For spectrum syntheses, we have selected a set of best MgH lines that are significant MgH contributors and are free or nearly free from other blends (see Table \ref{table:A.2}). The dissociation constant of MgH ($D_{0}$ = 1.34 eV) was sourced from the study of \cite{2013ApJS..207...26H}. The solar isotopic ratio for magnesium, $^{24}$Mg:$^{25}$Mg:$^{26}$Mg = 78.965:10.011:11.025, was adopted from \cite{2021A&A...653A.141A}. The LEP and $\log gf$ values for the selected lines are taken from the Kurucz database. Chris Sneden\footnote{\url{https://www.as.utexas.edu/~chris/lab.html}} has also reported fairly similar $\log gf$ and LEP values for $^{24}$MgH lines, along with the wavelengths of corresponding $^{25}$MgH and $^{26}$MgH lines, generated from the data published by \cite{2013ApJS..207...26H}.

\par To verify the adopted $gf$ values, we have independently calculated the oscillator strengths ($f$ values) using the relationship between $f$ and the Einstein $A$ coefficient (Eq. 11.12, \cite{gray_2021}). The Einstein $A$ coefficients were sourced from \cite{2013MNRAS.432.2043G}, as in Kurucz database, who calculated the $A$ values by combining the experimental potential curves and energy levels with high-quality $ab\: initio$ transition dipole moments using the relation defined by \cite{alma991043245695503276}. Note that, our independently determined $gf$ values are in excellent agreement with those adopted from Kurucz database. This validation attests to the reliability of the Kurucz database for $^{24}$MgH lines.

\par In the literature, we find that independent theoretical calculation by \cite{1979ApJ...231..637K} and \cite{2003ApJ...582.1059W} provide the band oscillator strength $f_{(0, 0)}$ for \mbox{MgH $A-X$ (0, 0)} molecular band. We note that, in the case of molecules, since the transitions are ro-vibronic (combined electronic, vibrational, rotational transitions) in nature, these transitions possess two different oscillator strengths: band ($f_{({\nu^\prime},{\nu^{\prime\prime}})}$) and rotational ($f_{({\nu^\prime}J^\prime, {\nu^{\prime\prime}}J^{\prime\prime})}$) oscillator strength. Oscillator strength defined for ro-vibronic transitions happening between two same or different vibrational levels is termed as band oscillator strength, whereas oscillator strength defined for ro-vibronic transitions happening between two different rotational levels belonging to two same or different vibrational levels is termed as rotational oscillator strength. In molecular transitions, the rotational oscillator strength is termed as the $f$ value, and it, combined with the statistical weight of the initial energy level ($g_{i}$), gives the commonly used $gf$ value \citep{2014ApJS..211....5R, 2014A&A...571A..47M}. $f_{({\nu^\prime},{\nu^{\prime\prime}})}$ is related to $f_{({\nu^\prime}J^\prime, {\nu^{\prime\prime}}J^{\prime\prime})}$ as follows \citep{2003ApJ...582.1059W}:
\begin{equation} \label{equ3}
    f_{({\nu^\prime},{\nu^{\prime\prime}})} = \frac{g_{{J^\prime}{J^{\prime\prime}}}}{{S_{J^\prime} ({J^{\prime \prime}})}}\times f_{({\nu^\prime}J^\prime, {\nu^{\prime\prime}}J^{\prime\prime})}
\end{equation}
where $S_{J^\prime} ({J^{\prime \prime}})$ is defined as the Höln-London factor. For the \mbox{MgH $A-X$ (0, 0)} band, the required $S_{J^\prime} (J^{\prime \prime})$ values for the P, Q and R molecular branches, defined by \cite{1974ApJS...27....1W}, are:
\begin{equation}
    S_{J^\prime} (J^{\prime \prime}) = 
    \left\{
	\begin{array}{lll}
		\frac{(J^{\prime\prime}\: -\: 1)}{2},  & J^\prime\: =\:           J^{\prime\prime}\: -\: 1 & {\rm (P-branch)} \\[0.1cm]
            \frac{(2J^{\prime\prime}\: +\: 1)}{2},  & J^\prime\: =\: J^{\prime\prime} & {\rm (Q-branch)} \\[0.1cm]
		\frac{(J^{\prime\prime}\: +\: 2)}{2},  & J^\prime\: =\:            J^{\prime\prime}\: +\: 1 & {\rm (R-branch)}
	\end{array}
    \right.
\end{equation}
Incorporating the $S_{J^\prime} ({J^{\prime \prime}})$ values from the above-mentioned relation, band oscillator strength ($f_{(0, 0)}$) of the \mbox{MgH $A-X$ (0, 0)} band was determined from our calculated rotational oscillator strengths, \textit{i.e.}, the $f$ values. Hence, our independently determined $f$ values above are actually the rotational oscillator strengths.

\par From Eq. \ref{equ3}, we determine an average band oscillator strength $f_{(0, 0)} = 0.1601$ and that is found to be in excellent agreement with independent theoretical calculations by \cite{1979ApJ...231..637K} and \cite{2003ApJ...582.1059W} ($f_{(0, 0)} = 0.161$).  \cite{10.1063/1.1675084}, using Hartree-Fock wave functions, derived a value of 0.250 for $f_{(0, 0)}$. Using the multi-configuration wavefunctions of \cite{10.1063/1.1673619}, \cite{10.1063/1.1674728} calculated $f_{(0, 0)} = 0.192$.

\par However, a significant lower value of $f_{(0, 0)} = 0.055$ was determined by \cite{1971MNRAS.154..265L}. To determine $f_{(0, 0)}$, \cite{1971MNRAS.154..265L} fit a straight line to the observations in a standard plot of log ($W_{\lambda}/S_{J^\prime} ({J^{\prime \prime}})$) versus $E_{J^{''}}$ ($W_{\lambda}$: Equivalent width and $E_{J^{''}}$: LEP) for a number of $^{24}$MgH lines present in the solar photosphere; the straight line fit to the plot is a model atmosphere prediction assuming LTE. Similarly, \cite{1973A&A....27...29G} empirically determined $f_{(0, 0)} = 0.035$ in order to get the best agreement between the predicted and observed solar equivalent widths of $^{24}$MgH lines. These discrepancies can be collectively attributed to the adopted solar magnesium abundance, the dissociation constant and other uncertainties arising from the adopted solar model atmosphere.

\par After successfully verifying the rotational as well as the band oscillator strengths of the adopted solar $^{24}$MgH molecular lines from the \mbox{$A-X$ (0, 0)} band, an abundance analysis for magnesium was performed using spectrum synthesis. A set of these MgH lines (see Table. \ref{table:A.2}) was synthesized for eight different He/H ratios as mentioned above. Synthesis of \mbox{MgH $A-X$ (0, 0)} R$_{1}$13 line is shown in Figure \ref{fig:synth}(b) as an example.

\vspace{0.5cm}

\textbf{CH:} For spectrum synthesis, the solar CH molecular lines of the \mbox{$A-X$ (0, 0)} and (1, 1) molecular bands are adopted from \cite{2021A&A...656A.113A}. Three more \mbox{CH $A-X$ (0, 0)} lines at 4218.724{\AA}, 4248.939{\AA}, and 4356.361{\AA}, listed by \cite{2005A&A...431..693A}, were also added to our adopted linelist from \cite{2021A&A...656A.113A}. The lower excitation potential and the transition probability values for these individual lines are from \cite{2014A&A...571A..47M}. The dissociation constant for CH ($D_{0}$ = 3.465 eV) was sourced from \cite{Huber1979}. The solar isotopic ratio for carbon, $^{12}$C:$^{13}$C = 98.893:1.107, was adopted from \cite{2021A&A...653A.141A} and the wavelengths of corresponding $^{13}$CH lines were taken from \cite{2014A&A...571A..47M}. \cite{2021A&A...656A.113A} used the LEP and $\log gf$ values sourced from 
\cite{2014A&A...571A..47M}, while \cite{2005A&A...431..693A} adopted these values from \cite{FOLOMEG1987562}. We found that the values listed in both sources are in good agreement with each other. Hence, we adopted the values provided by \cite{2014A&A...571A..47M} as it is the most recent source. Syntheses of \mbox{CH $A-X$ (0, 0)} R$_{2e}$10 and R$_{1f}$10 lines are shown in Figure \ref{fig:synth}(c) as examples.

\vspace{0.5cm}

\textbf{C$_{2}$ Swan:} The solar C$_{2}$ Swan molecular lines of the (0, 0) molecular band are from \cite{2005A&A...431..693A}. The dissociation constant value ($D_{0}$ = 6.297 eV) is from \cite{URDAHL1991425}. The solar carbon isotopic ratio, adopted for CH molecular lines, was also applied to the C$_{2}$ Swan molecular lines. The wavelengths of corresponding $^{12}$C$^{13}$C lines were taken from \cite{2013JQSRT.124...11B}. For the analysis, we have considered three different sources that provide the $\log gf$ values for the \mbox{C$_{2}$ Swan (0, 0)} transitions. These are \cite{1991A&A...242..488G}, \cite{2012ApJ...747..102H}, and \cite{2013JQSRT.124...11B}. \cite{1991A&A...242..488G} provide values from measurements of the $d^3\Pi_{g}$ molecular state's radiative lifetime. \cite{2012ApJ...747..102H} provide $gf$-values that are from the theoretical band oscillator strengths computed by \cite{10.1063/1.2806988}. \cite{2013JQSRT.124...11B}'s study is the latest in the literature and is based on $ab\: initio$ calculation of the transition dipole moment function. These three sources also provide the transition's lower excitation potential but note that its $gf$-value as well as the LEP differs from one source to another.

\par An abundance analysis of carbon was conducted by synthesizing the C$_{2}$ Swan transitions using the standard ATLAS12 solar model atmosphere for He/H ratio 0.1. These three sources provide three different pairs of (LEP, $\log gf$). The best fit to the observed C$_{2}$ Swan transition hence provides the carbon abundance. Note that the carbon abundances derived from these three different sources are in good agreement within 0.05 dex. In this study, we finally adopt \cite{2013JQSRT.124...11B}'s values for the subsequent abundance analysis of carbon. Synthesis of \mbox{C$_{2}$ Swan (0, 0)} R$_{1}$11 line is shown in Figure \ref{fig:synth}(d) as an example.

\subsubsection{Atomic lines}\label{sec:MgI}
\textbf{Mg\,{\sc i}:} An equivalent width analyses was conducted for the measured equivalent widths of neutral magnesium (Mg\,{\sc i}) lines. The atomic data for these transitions for example, the line's wavelength, the LEP, and the transition probability \textit{i.e.}, the $\log gf$ value are from two sources; \cite{2015A&A...573A..25S} and \cite{2021A&A...653A.141A}. Both \cite{2015A&A...573A..25S} and \cite{2021A&A...653A.141A} provide the measured equivalent widths of these observed transitions in Sun, and for the common lines these are in excellent agreement with our measurements using the FTS solar spectrum. \cite{2021A&A...653A.141A}'s list includes two additional lines of Mg\,{\sc i} at 8712.689{\AA} and 8717.825{\AA}, and adopts the recent $\log-gf$ values from \cite{2017A&A...598A.102P}. 

\par Our analyses of both these lists confirm that the derived Mg abundances are in excellent agreement, however, the line-to-line scatter is larger for \cite{2015A&A...573A..25S}'s list. Hence, we adopted \cite{2021A&A...653A.141A}'s list and Mg abundances were derived for 8 different He/H ratios as mentioned above. In this study, one more Mg\,{\sc i} line at 5711.088{\AA} was added to \cite{2021A&A...653A.141A}'s list as this line was found to be clean and without blends; the $\log-gf$ value is from \cite{2017A&A...598A.102P} and the LEP is from the {\it NIST}\footnote{\url{https://www.nist.gov/pml/atomic-spectra-database}} Atomic Spectra Database.

\vspace{0.5cm}

\textbf{C\,{\sc i} and [C\,{\sc i}]:} An abundance analyses was conducted for the neutral carbon lines, both permitted (C\,{\sc i}) and forbidden [C\,{\sc i}] lines were considered. The measured equivalent widths including the atomic data are from \cite{2019A&A...624A.111A}. Note that, the infra-red (IR) lines were excluded from our LTE analysis as these lines exhibit severe departures from LTE \citep{2005A&A...431..693A}.

\par The adopted equivalent widths are in excellent agreement with our measurements, except for the forbidden carbon [C\,{\sc i}] line. Our measured equivalent width for the [C\,{\sc i}] line is however, close to \cite{1978MNRAS.182..249L}'s measured value of 6.5 m{\AA}. Hence, we adopt \cite{1978MNRAS.182..249L}'s measurement over \cite{2019A&A...624A.111A}, that is 4.7 m{\AA}, for the [C\,{\sc i}] line.

\par
\begin{figure*}
    \centering
    \includegraphics[scale=0.335]{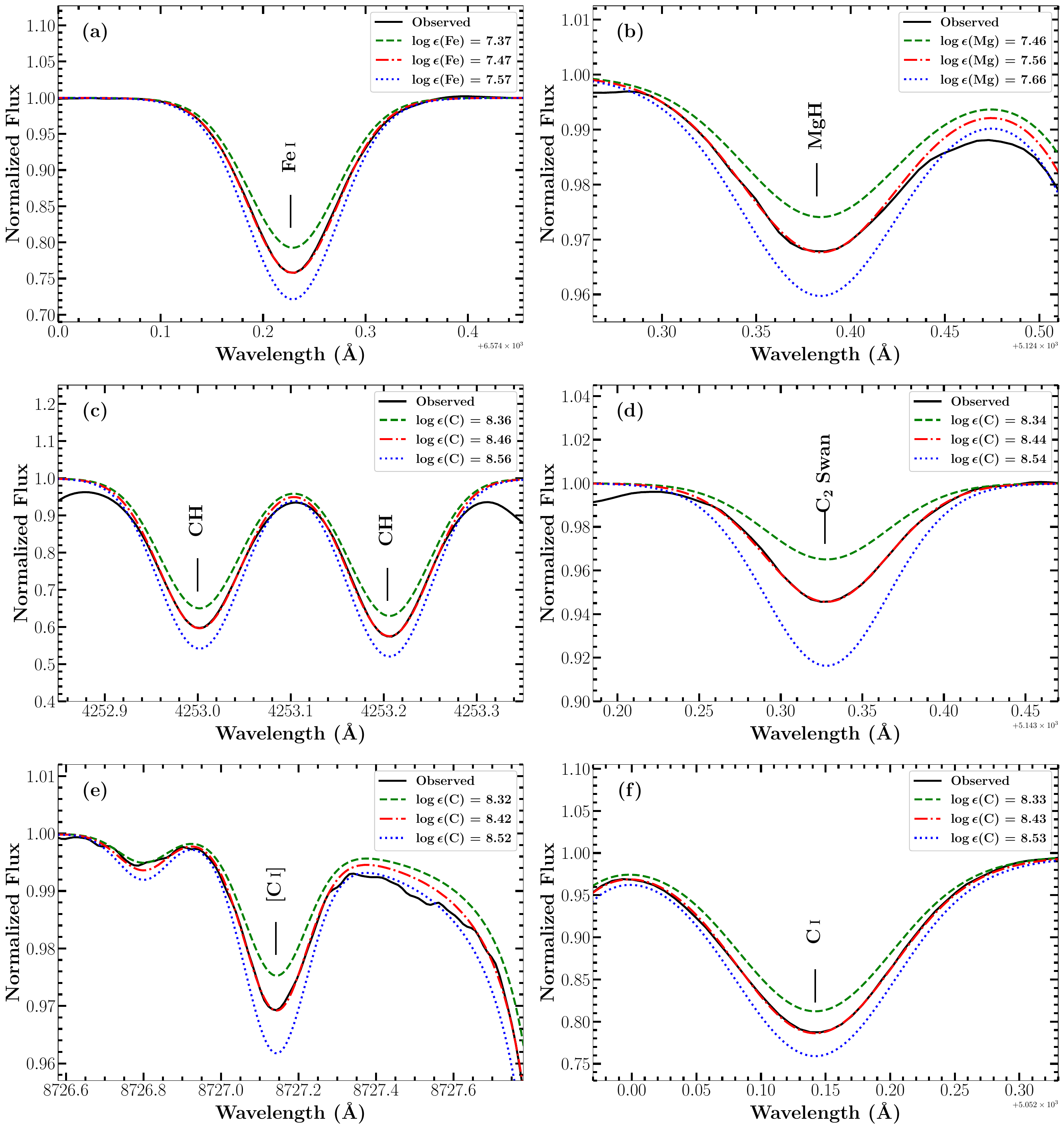}
    \caption{Syntheses of spectral lines of various species for He/H = 0.1. (a) Fe\,{\sc i} line at 6574.229{\AA}. (b) \mbox{MgH $A-X$ (0, 0)} R$_{1}$13 line. (c) \mbox{CH $A-X$ (0, 0)} R$_{2e}$10 and R$_{1f}$10 lines. (d) \mbox{C$_{2}$ Swan (0, 0)} R$_{1}$11 line. (e) The forbidden [C\,{\sc i}] line. (f) C\,{\sc i} line at 5052.149{\AA}.}
    \label{fig:synth}
\end{figure*}

 Finally, carbon abundances were derived from all these line transitions for 8 different He/H ratios. We have derived the abundances using the equivalent width analysis as well as the spectrum synthesis. The derived abundances from both these methods are in excellent agreement. Nevertheless, we report the derived abundances obtained from spectrum synthesis. Syntheses of the forbidden [C\,{\sc i}] line and the permitted C\,{\sc i} line at 5052.149{\AA} are shown in Figure \ref{fig:synth}(e)
and \ref{fig:synth}(f), respectively, as examples.

\begin{table*}
\centering
\caption{Abundance of Mg and C obtained from their key species for different He/H models.}
\label{table:1}
\begin{tabular}{cccccccc}\toprule\toprule
\multirow{2}{*}{He/H} &\multicolumn{2}{c}{Mg} &\multicolumn{5}{c}{C}\\\cmidrule(r){2-3}\cmidrule(l){4-8}
&$\log \epsilon$(Mg)$_{\rm Mg\,\textsc{i}}$ &$\log \epsilon$(Mg)$_{\rm MgH}$ &$\log \epsilon$(C)$_{\rm C\,\textsc{i}}$ &$\log \epsilon$(C)$_{\rm [C\,\textsc{i}]}$ &$\log \epsilon$(C)$_{\rm C\,\textsc{i}\, +\, [C\,\textsc{i}]}$ &$\log \epsilon$(C)$_{\rm CH\, (\textit{A-X})}$ &$\log \epsilon$(C)$_{\rm C_{2}\, Swan}$ \\\midrule
0.075 &7.60 $\pm$ 0.06 &7.52 $\pm$ 0.05 &8.48 $\pm$ 0.03 &8.44 &8.47 $\pm$ 0.03 &8.43 $\pm$ 0.02 &8.43 $\pm$ 0.02 \\
0.085 &7.57 $\pm$ 0.06 &7.53 $\pm$ 0.05 &8.45 $\pm$ 0.04 &8.43 &8.45 $\pm$ 0.04 &8.44 $\pm$ 0.02 &8.43 $\pm$ 0.02 \\
0.100 &7.55 $\pm$ 0.06 &7.54 $\pm$ 0.05 &8.42 $\pm$ 0.04 &8.42 &8.42 $\pm$ 0.04 &8.44 $\pm$ 0.02 &8.43 $\pm$ 0.02 \\
0.125 &7.53 $\pm$ 0.06 &7.56 $\pm$ 0.05 &8.38 $\pm$ 0.04 &8.41 &8.38 $\pm$ 0.04 &8.46 $\pm$ 0.02 &8.43 $\pm$ 0.02 \\
0.135 &7.51 $\pm$ 0.05 &7.58 $\pm$ 0.05 &8.34 $\pm$ 0.05 &8.40 &8.35 $\pm$ 0.05 &8.47 $\pm$ 0.02 &8.43 $\pm$ 0.02 \\
0.150 &7.50 $\pm$ 0.05 &7.59 $\pm$ 0.05 &8.30 $\pm$ 0.05 &8.40 &8.32 $\pm$ 0.06 &8.48 $\pm$ 0.02 &8.43 $\pm$ 0.02 \\
0.175 &7.47 $\pm$ 0.05 &7.62 $\pm$ 0.05 &8.25 $\pm$ 0.05 &8.39 &8.27 $\pm$ 0.07 &8.50 $\pm$ 0.02 &8.43 $\pm$ 0.02 \\
0.200 &7.45 $\pm$ 0.05 &7.64 $\pm$ 0.05 &8.21 $\pm$ 0.05 &8.37 &8.23 $\pm$ 0.08 &8.52 $\pm$ 0.02 &8.43 $\pm$ 0.02 \\
\bottomrule
\end{tabular}
\end{table*}

\subsection{Determination of solar He/H ratio} \label{sec:hebyh}
Table \ref{table:1} illustrates the abundance of magnesium obtained from Mg\,{\sc i} atomic lines as well as from MgH molecular lines. Eight sets of Mg abundances are listed for the adopted eight different He/H ratios (see Table \ref{table:1}). Similarly, Table \ref{table:1} also illustrates eight sets of carbon abundances derived from neutral carbon transitions, both permitted and forbidden, and from molecular lines of CH and C$_{2}$ Swan.

\par
\begin{figure*}
    \centering
    \includegraphics[scale=0.375]{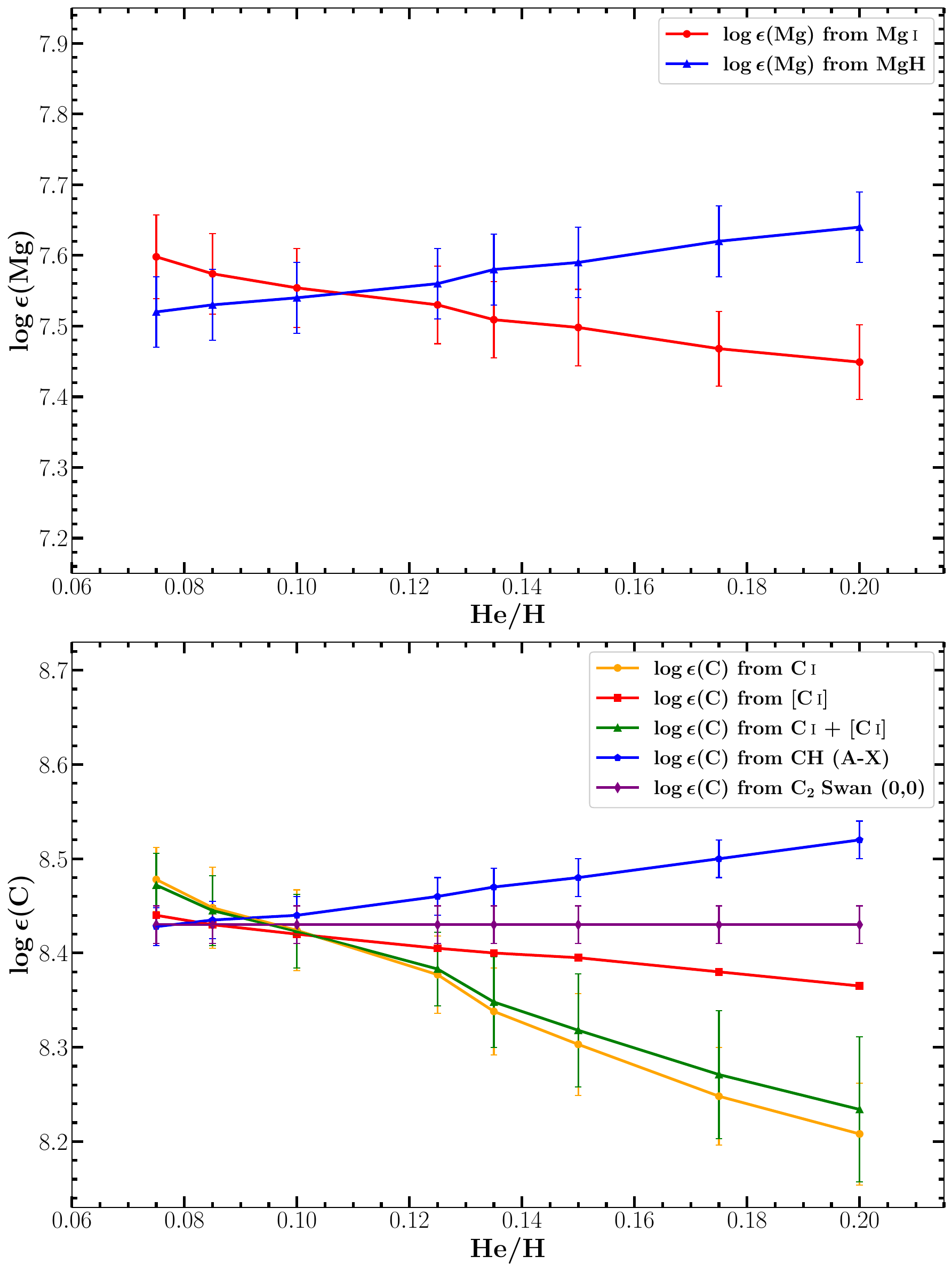}
    \caption{Abundance of magnesium obtained from Mg\,{\sc i} and MgH lines for different He/H ratios (top panel). Abundance of carbon obtained from C\,{\sc i}, CH, and C$_{2}$ Swan lines for different He/H ratios (bottom panel).} \label{fig:mg_c}
\end{figure*}

 Mg and C abundances, derived from their observed atomic and molecular absorptions, versus the adopted model's He/H ratios are shown in Figures, \ref{fig:mg_c}(top panel) and \ref{fig:mg_c}(bottom panel), respectively. An examination of the Figures, \ref{fig:mg_c}(top panel) and \ref{fig:mg_c}(bottom panel), suggest that the derived Mg and C abundances depend on the adopted model's He/H ratio except for the derived C abundance from C$_{2}$ Swan transitions. It is worth noting that for a higher He/H ratio, the derived Mg and C abundances from their observed atomic lines are lower than those derived for a lower He/H ratio. However, the derived Mg and C abundances from their respective hydrides exhibit an inverse trend (See Figures, \ref{fig:mg_c}(top panel) and \ref{fig:mg_c}(bottom panel)). These trends are as expected due to the adopted model's He/H ratio; decreasing the abundance of hydrogen or increasing the abundance of helium, \textit{i.e.}, increasing the He/H ratio, results in a decrease in continuous opacity per gram \citep{2011ApJ...737L...7S} along with a decrease in the availability of hydrogen atoms to form metal hydrides. Therefore, for the same observed strength of the atomic line, the elemental abundance must decrease \citep{2020ApJ...897...32H}. But for a metal hydride line, a combined effect of the reduced continuum absorption and the line's reduced absorption strength demands an increased metal abundance to fit the same observed line strength.

\par In principle, the abundances of magnesium and carbon obtained from their respective atomic and molecular lines must return the same abundances within the measured uncertainty. Here, we note that the rms errors in abundances due to line-to-line scatter dominate over the other measurement errors. For example, the uncertainty in measuring the equivalent width or the errors associated with the parameters involving the spectrum syntheses are not very significant.  We infer the He/H ratios of 0.108 $^{+ 0.051}_{- 0.046}$ and 0.091 $^{+ 0.019}_{- 0.014}$ as the best-determined values from Mg and C abundance analyses, respectively (see Figures, \ref{fig:mg_c}(top panel) and \ref{fig:mg_c}(bottom panel)). The uncertainties in the derived He/H ratios are translated from the rms uncertainties in abundances associated with the atomic and molecular hydride features of Mg and C. For the adopted range in the He/H values (see Figures, \ref{fig:mg_c}(top panel) and \ref{fig:mg_c}(bottom panel)), note the divergence in the derived Mg and C abundances from their respective features and the associated abundance uncertainties. For this study, we adopted models with a range in their He/H values: 0.075 $\leq$ He/H $\leq$ 0.200; more weight is given to the abundance analyses of C than that of Mg due to the lower uncertainties in the derived C abundances.

\section{Conclusion and Discussions} \label{candd}
\par In Table \ref{tab:compare}, we compare the derived abundances of the key species using three different solar 1D LTE model atmospheres: ATLAS12 (this study), MARCS\footnote{The theoretical hydrostatic model computed using the MARCS code \citep{2008A&A...486..951G}.}, and HM\footnote{The semi-empirical Holweger-Müller model \citep{1974SoPh...39...19H}, upgraded from the \cite{1967ZA.....65..365H} version using updated equation of state and continuous opacities \citep{2009ARA&A..47..481A}.}. The derived abundances in Table \ref{tab:compare} are for the solar model with He/H = 0.1. Table \ref{tab:compare} clearly demonstrates that the derived abundances in this study are in excellent agreement with that of \cite{2021A&A...653A.141A}.

\begin{table*}
\centering
\caption{Comparison of abundances derived using 1D LTE Model atmosphere for He/H = 0.1: ATLAS12 (this work) and \cite{2021A&A...653A.141A}.}\label{tab:compare}
\begin{tabular}{ccccc}\toprule\toprule
\multirow{2}{*}{Abundance} &\multirow{2}{*}{Species} &This work &\multicolumn{2}{c}{Asplund \textit{et al.} (2021)} \\\cmidrule{4-5}
&&ATLAS12 &MARCS &HM \\\midrule
\multirow{2}{*}{$\log \epsilon$(Fe)} &Fe\,{\sc i} &7.43 $\pm$ 0.05 &7.41 $\pm$ 0.04 &7.48 $\pm$ 0.05 \\
&Fe\,{\sc ii} &7.43 $\pm$ 0.07 &7.38 $\pm$ 0.04 &7.43 $\pm$ 0.03 \\\hline
\multirow{2}{*}{$\log \epsilon$(Mg)} &Mg\,{\sc i} &7.55 $\pm$ 0.06 &7.52 $\pm$ 0.02 &7.57 $\pm$ 0.03 \\
&MgH &7.54 $\pm$ 0.05 &\nodata &\nodata \\\hline
\multirow{4}{*}{$\log \epsilon$(C)} &[C\,{\sc i}] &8.42 &8.42 &8.43 \\
&C\,{\sc i} &8.42 $\pm$ 0.04 &8.46 $\pm$ 0.04 &8.50 $\pm$ 0.04 \\
&CH ($A-X$) &8.44 $\pm$ 0.02 &8.40 $\pm$ 0.05 &8.56 $\pm$ 0.05 \\
&C$_{2}$ Swan &8.43 $\pm$ 0.02 &8.43 $\pm$ 0.03 &8.52 $\pm$ 0.03 \\
\bottomrule
\end{tabular}
\end{table*}

\cite{2003ApJ...591.1220L} suggested present-day solar helium abundance of $\log \epsilon$(He) = 10.899 $\pm$ 0.005 from averaging helium abundance values obtained from various helioseismic studies over the years. \cite{2004ApJ...606L..85B} have also derived the solar helium mass fraction, $Y_{\odot}$, as 0.2485 $\pm$ 0.0034 using helioseismology, which corresponds to a He/H ratio of 0.085 or $\log \epsilon$(He) = 10.93 $\pm$ 0.01. With the improved SAHA-S3 equation of state, \cite{2014MNRAS.441.3296V}, derived a range for the solar helium mass fraction, $Y_{\odot}$, as 0.240-0.255.

\par \cite{2021A&A...653A.141A} has reported $Y_{\odot}$ = 0.2423 $\pm$ 0.0054 by taking the mean of \cite{2004ApJ...606L..85B} and \cite{2014MNRAS.441.3296V}. This corresponds to a He/H ratio of 0.082 or $\log \epsilon$(He) = 10.914 $\pm$ 0.013.

\par Our results determined from the observed absorptions of Mg\,{\sc i} and MgH and that of C\,{\sc i} and CH, are consistent; He/H = 0.108 $^{+ 0.051}_{- 0.046}$ and 0.091 $^{+ 0.019}_{- 0.014}$ from the abundance analyses of Mg and C, respectively. Our derived He/H ratios are in fair agreement with the result obtained through various helioseismological studies, signifying the reliability and accuracy of our novel technique in determining the solar helium-to-hydrogen ratio. This study also confirms that the widely assumed and adopted (He/H)$_{\odot}$ = 0.1 is in fair agreement with our measurements. More reliable values should, in principle, come from 3D model atmospheres with full non-LTE calculations.

\par Using our derived He/H ratio (0.091 $^{+ 0.019}_{- 0.014}$) and \cite{2021A&A...653A.141A}'s $(Z/X)_{\odot}$ value, we have determined the solar mass fraction as $X_{\odot}$ = 0.7232 $^{+ 0.0305}_{- 0.0377}$, \mbox{$Y_{\odot}$ = 0.2633 $^{+ 0.0384}_{- 0.0311}$}, and $Z_{\odot}$ = 0.0135 $^{+ 0.0006}_{- 0.0007}$. These values strongly constrain the modeling of the structure and evolution of the Sun. It will be noteworthy to see whether the standard stellar evolution model constructed with our deduced values of $X_{\odot}$, $Y_{\odot}$, and $Z_{\odot}$ can reproduce the present solar luminosity $L_{\odot}$ at the present solar age $t_{\odot}$.
\section{Acknowledgments}

We are grateful to the referee for a fine and constructive report. GP thanks DST/SERB for their support through CRG grant CRG/2021/000108. BPH acknowledges and thanks the Women Scientist Scheme-A (WOS-A), Department of Science and Technology (DST), India, for support through Grant: DST/WOS-A/PM-1/2020. SM acknowledges the support of the Indian Institute of Astrophysics, Bangalore for conducting this research.

\appendix

\section{Linelists of Mg\,\textsc{i}, MgH, C\,\textsc{i}, CH and C$_{2}$ Swan spectral lines}
\centering
\setcounter{figure}{0} \renewcommand{\thefigure}{A.\arabic{figure}}
\setcounter{table}{0} \renewcommand{\thetable}{A.\arabic{table}}

\begin{table*}[!htp]
\centering
\caption{Abundance of Mg derived from Mg\,{\sc i} atomic lines for He/H = 0.1.}
\label{table:A.1}
\begin{tabular}{ccccc}\toprule\toprule
$\lambda$ &LEP &$\log gf$ &EW &$\log \epsilon$(Mg) \\
({\AA}) &(eV) & &(m{\AA})& \\\midrule
5711.088 &4.346 &-1.742 &113.5 &7.51 \\
6318.716 &5.108 &-2.020 &41.3 &7.54 \\
6319.236 &5.108 &-2.242 &26.0 &7.50 \\
8712.689 &5.932 &-1.152 &68.0 &7.57 \\
8717.825 &5.933 &-0.930 &100.0 &7.59 \\
8923.569 &5.394 &-1.679 &63.3 &7.64 \\
9429.814 &5.932 &-1.306 &47.1 &7.48 \\
9983.200 &5.932 &-2.177 &10.0 &7.55 \\
10312.531 &6.118 &-1.718 &18.3 &7.51 \\
11522.240 &6.118 &-1.913 &21.0 &7.67 \\
12417.937 &5.932 &-1.662 &44.8 &7.55 \\
12423.029 &5.932 &-1.185 &97.0 &7.55 \\
\midrule
\multicolumn{5}{c}{avg. $\log \epsilon$(Mg) = 7.55 $\pm$ 0.06} \\
\bottomrule
\end{tabular}
\end{table*}

\begin{table*}[!htp]
\centering
\caption{Abundance of Mg derived from \mbox{MgH $A-X$ (0, 0)} molecular lines for He/H = 0.1.}
\label{table:A.2}
\begin{tabular}{ccccc}\toprule\toprule
$\lambda$ &Branch &LEP &$\log gf$ &$\log \epsilon$(Mg) \\
({\AA}) & &(eV) & & \\\midrule
5124.411 &$R_{1}$13 &0.128 &0.105 &7.56 \\
5153.680 &$Q_{1}$18 &0.238 &0.486 &7.50 \\
5198.326 &$P_{2}$26 &0.478 &0.274 &7.55 \\
5201.636 &$P_{1}$6 &0.030 &-0.346 &7.48 \\
5202.985 &$P_{1}$24 &0.411 &0.261 &7.47 \\
5207.083 &$P_{2}$21 &0.319 &0.183 &7.60 \\
5209.590 &$P_{1}$19 &0.264 &0.165 &7.58 \\
\midrule
\multicolumn{5}{c}{avg. $\log \epsilon$(Mg) = 7.54 $\pm$ 0.05} \\
\bottomrule
\end{tabular}
\end{table*}

\newpage
\begin{table*}[!htp]
\centering
\caption{Abundance of C derived from C\,{\sc i} atomic lines for He/H = 0.1.}
\label{table:A.3}
\begin{threeparttable}
\begin{tabular}{ccccc}\toprule\toprule
$\lambda$ &LEP &$\log gf$ &$\log \epsilon$(C) \\
({\AA}) &(eV) & & \\\midrule
8727.126$^*$ &1.264 &-8.165 &8.42 \\\midrule
5052.149 &7.685 &-1.303 &8.43 \\
5380.331 &7.685 &-1.616 &8.48 \\
6587.608 &8.537 &-1.003 &8.38 \\
7111.475 &8.640 &-1.085 &8.37 \\
7113.180 &8.647 &-0.773 &8.46 \\
\midrule
\multicolumn{4}{c}{avg. $\log \epsilon$(C) = 8.42 $\pm$ 0.04} \\
\bottomrule
\end{tabular}
\begin{tablenotes}
    \item [*]
Forbidden line.
\end{tablenotes}
\end{threeparttable}
\end{table*}

\begin{table*}[!htp]
\centering
\caption{Abundance of C derived from CH molecular lines for He/H = 0.1.}
\label{table:A.4}
\begin{threeparttable}
\begin{tabular}{cccccc}\toprule\toprule
$\lambda$ &Band &Branch &LEP &$\log gf$ &$\log \epsilon$(C) \\
({\AA}) & & &(eV) & & \\\midrule
4218.723 &$A-X$(0, 0) &$R_{2e}$15 &0.411 &-1.008 &8.41\\
4248.945 &$A-X$(0, 0) &$R_{1f}$15 &0.189 &-1.431 &8.47\\
4253.003 &$A-X$(1, 1) &$R_{2e}$10 &0.523 &-1.506 &8.46 \\
4253.209 &$A-X$(1, 1) &$R_{1f}$10 &0.523 &-1.471 &8.46 \\
4255.252 &$A-X$(0, 0) &$R_{1f}$9 &0.157 &-1.455 &8.43 \\
4263.976 &$A-X$(1, 1) &$R_{2e}$8 &0.460 &-1.575 &8.43 \\
4274.186 &$A-X$(0, 0) &$R_{1e}$6 &0.074 &-1.563 &8.46 \\
4356.375 &$A-X$(0, 0) &$P_{2f}$9 &0.155 &-1.846 &8.43\\
4356.600 &$A-X$(0, 0) &$P_{1e}$9 &0.157 &-1.793 &8.44\\
\midrule
\multicolumn{6}{c}{avg. $\log \epsilon$(C) = 8.44 $\pm$ 0.02} \\
\bottomrule
\end{tabular}
\end{threeparttable}
\end{table*}

\begin{table*}[!htp]
\centering
\caption{Abundance of C derived from \mbox{C$_{2}$ Swan (0, 0)} molecular lines for He/H = 0.1.}
\label{table:A.5}
\begin{threeparttable}
\begin{tabular}{ccccc} \toprule\toprule
$\lambda$ &Branch &LEP &$\log gf$ &$\log \epsilon$(C) \\
({\AA}) & &(eV) & & \\\midrule
5033.700 &$R_{3}$50 &0.508 &0.193 &8.44 \\
5073.600 &$R_{3}$39 &0.312 &0.082 &8.45 \\
5109.300 &$R_{3}$27 &0.152 &-0.082 &8.41 \\
5132.500 &$R_{2}$17 &0.062 &-0.262 &8.43 \\
5136.600 &$R_{3}$15 &0.049 &-0.341 &8.38 \\
5140.400 &$R_{3}$13 &0.037 &-0.404 &8.45 \\
5143.300 &$R_{1}$11 &0.026 &-0.409 &8.44 \\
5144.900 &$R_{1}$10 &0.022 &-0.445 &8.43 \\
\midrule
\multicolumn{5}{c}{avg. $\log \epsilon$(C) = 8.43 $\pm$ 0.02} \\
\bottomrule
\end{tabular}
\end{threeparttable}
\end{table*}

\newpage


\end{document}